\documentclass[12pt,a4paper]{article}

\usepackage{amsmath,amsthm,amsfonts}

\usepackage{color,amsmath,amssymb,amsfonts}
\usepackage{url,lscape}
\usepackage[authoryear]{natbib}
\usepackage{multirow}

\RequirePackage[dvips]{graphicx}
\usepackage{epsf,epsfig,subfigure,psfrag,graphicx,afterpage}
\newcommand{\bftheta  }{\hbox{\boldmath$\theta$}}

\begin{document}

\title{Stochastic Volatily Models using Hamiltonian Monte Carlo
  Methods and Stan}

\author{
  David S. Dias$^{\rm a}$
  Ricardo S. Ehlers$^{\rm b}$\thanks{Corresponding author. Email: ehlers@icmc.usp.br}
  \vspace{6pt}\\
  $^{\rm a}${\em ICMC/USP and UFSCar, Brazil}
  $^{\rm b}${\em University of S\~ao Paulo, S\~ao Carlos, Brazil}
}

\date{}

\maketitle

\begin{abstract}
  
This paper presents a study using the Bayesian approach in stochastic volatility models for modeling
financial time series, using Hamiltonian Monte Carlo methods (HMC). We propose the use
of other distributions for the errors in the observation equation of stochastic volatiliy models, besides the
Gaussian distribution, to address problems as heavy tails and asymmetry in the returns. Moreover,
we use recently developed information criteria WAIC and LOO that approximate the cross-validation methodology, 
to perform the selection of models. Throughout this work, we study the
quality of the HMC methods through examples, simulation studies and applications to real datasets.
\vskip .3cm


\end{abstract}

\section{Introduction}

Stochastic volatility (SV) models have been around for decades now and succesfully applied to study the volatility which is characteristic in financial markets.
A number of estimation methods have been proposed to estimate these models, but 
Markov Chain Monte Carlo (MCMC) are usually considered one of the most efficient methods. 
Great advances have been made recently with the Hamiltonian Monte Carlo
algorithms (HMC, see for example \cite{nea2011}) for the estimation of latent variable models. The practitioner, usually interested in fast answers, has benefited from these recent methodological developements and most importantly from availability of computational programs.

The computations in this paper were implemented
using open-source statistical software. In particular, we used the
{\tt R} environment (\cite{r10}) and recently released {\tt R} packages
which can efficiently estimate stochastic volatility (SV) models. 
The {\tt rstan} package is
an interface to the open-source Bayesian software Stan
(\cite{rstan-software:2016}) and the {\tt stochvol} package was proposed
by \cite{kastner2016Stochvol} which jointly samples all
instantaneous volatilities ``all without a loop'' (AWOL), a technique
discussed in more detail in \cite{mccausland2011simulation} and \cite{kastner2014ancillarity}. 
The {\tt stochvol} package was designed to reduce serial correlations of the MCMC
draws significantly and uses the auxiliary finite mixture
approximation of the errors as described in \cite{kim98} and
\cite{omori-etal}. As such, it is crafted to provide efficient
estimation of SV models with normal errors.

Stan on the other hand is actually a language designed for Bayesian
analysis with continuous parameter spaces and can be run from 
{\tt R}. Also, Stan uses Hamiltonian Monte Carlo (HMC) methods coupled
with the no-U-turn sampler (NUTS) which are designed to improve
speed, stability and scalability compared to standard MCMC as
Metropolis-Hastings and the Gibbs sampler. In particular, HMC methods
are also designed to reduce serial correlations of the MCMC draws and
usually the chains reach the stationary distribution with fewer
iterations.

In this paper, we first compare the {\tt rstan} and {\tt stochvol}
packages to estimate SV model with normal errors. Being more flexible than {\tt stochvol} in terms of
likelihood and prior specifications checking whether Stan is at least
as efficient provides a useful information for the applied researcher.
We then move to
explore Stan facilities to specify more flexible SV models with heavy
tailed distributions for the errors. These models are subsequently compared in
terms of recently proposed information criteria, namely the
\textit{Watanabe information criterion} (WAIC, \cite{waic}) and the 
\textit{Approximate Leave-One-Out Cross-Validation} (LOO, \cite{vehtari-etal}).

The rest of the paper is structured as follows. In Section
\ref{methods} SV models are briefly reviewed and the prior
distributions are described. The methodology for estimating and
comparing models is also described here. These methods are assessed
through a simulation study in Section
\ref{simulations} and through the statistical analysis of
real time series in section \ref{applications}.
Some final
comments are given in Section \ref{conclusion}.

\section{Model Description and Methods} \label{methods}

The canonical form of the SV model as given in \cite{kim98} is defined as,
\begin{align} 
  r_t &= \beta\exp({h_t/2})\varepsilon_t, \label{modSV}    \\  
  h_t &= \mu + \phi(h_{t-1}-\mu) + \eta_t, \label{log_vol} \\   
  h_1&\sim\mathcal{N}\left(\mu,\frac{\sigma_\eta^2}{1-\phi^2}\right), \nonumber  
\end{align}

\noindent where $h_t$ is a stationary process describing the
log-volatility at time $t$, $\{\varepsilon_t\}$ is a sequence of
independent and identicaly distributed random variables with mean zero
and variance one $\{\eta_t\}$ is a sequence such that $\eta_t\sim\mathcal{N}(0,\sigma_\eta^2)$,
with $\varepsilon_t$ and $\eta_t$ uncorrelated for all $t$. Then $\mu$
is the level of log-volatilities and $|\phi|<1$ is the persistence
parameter. The parameter $\beta$ is playing the role of a scale factor.
 
In its original formulation introduced by \cite{taylor82},
$\varepsilon_t$ is assumed to follow a standard normal
distribution. However, several other error distributions have been
proposed as many empirical studies indicate that the canonical model
does not account for the amount of kurtosis usually observed in most
financial time series returns. Therefore, we also consider other
(heavy tailed) distributions for the error term in the observation
equation (\ref{modSV}) which are described below.

The Exponential
Power distribution (or generalized error distribution, GED)
with mean zero and variance one whose density function is given by,
\begin{equation}
p(\varepsilon_t) = 
\frac{\nu}{\lambda 2^{\big(1 + \frac{1}{\nu}\big)}\Gamma\big(\frac{1}{\nu}\big)}
\exp\left\{-\frac{1}{2}\Big|\frac{\varepsilon_t}{\lambda}\Big|^{\nu}\right\}.
\end{equation}

\noindent In this formulation, 
$\lambda^2 = 2^{-\frac{2}{\nu}}
\Gamma\left(\frac{1}{\nu}\right)/\Gamma\left(\frac{3}{\nu}\right)$
and $\nu>0$ is the shape parameter. It is not difficult to see that
the double exponential or Laplace
distribution is obtained for $\nu=1$ and the standard normal
distribution for $\nu=2$. Also, the kurtosis is given by
$\Gamma\left(\frac{1}{\nu}\right)\Gamma\left(\frac{5}{\nu}\right)
\Gamma\left(\frac{3}{\nu}\right)^2 - 3$ and then $0<\nu < 2$ leads to
heavy tail distributions.

The $t$-Student distribution with $\nu$ degrees of feedom and density
function given by,
\begin{equation}
p(\varepsilon_t) = \frac{1}{\sqrt{\pi(\nu-2)}}\frac{\Gamma\big(\frac{\nu+1}{2}\big)}{\Gamma\big(\frac{\nu}{2}\big)}\left\{1+\frac{\varepsilon_t^2}{\nu-2}\right\}^{-\frac{\nu+1}{2}}.
\end{equation}
when $\nu\rightarrow\infty$ this approaches the density of a standard
normal distribution.

Additionaly, we consider a standardized Skew-Normal distribution with
shape parameter $\nu$ and density function given by,
\begin{equation}
p(\varepsilon_t) = \frac{1}{\sqrt{2\pi}}\exp\left\{-\frac{\varepsilon_t^2}{2}\right\}\left\{1+\textrm{erf}\left[\nu\left(\frac{\varepsilon_t}{\sqrt{2}}\right)\right]\right\},
\end{equation}
where $\textrm{erf}(x)=\frac{2}{\sqrt\pi}\int_{0}^{2}e^{-t^2}dt$. This
distribution is skewed to the left for $\nu<0$ and right skewed for
$\nu>0$ thus taking into account another empirical evidence that many
returns present a slight skewness. The standard normal is a particular
case for $\nu=0$.

To complete the model specification under the Bayesian paradigm we
define independent
prior distributions for the parameters $\phi$ and
$\sigma^2_{\eta}$. As in \cite{kim98} the prior for $\phi$ is
specified by setting $\phi=2\phi^{*} -1$ where $\phi^{*} \sim
\mathcal{B}(\alpha; \beta)$ so that,

\begin{equation}\label{priorPhi}
\pi(\phi) = \frac{1}{2B(\alpha,\beta)}
\left(\frac{1+\phi}{2}\right)^{\alpha-1}\left(\frac{1-\phi}{2}\right)^{\beta-1},
\qquad \alpha,\beta > \frac{1}{2},
\end{equation}

\noindent where $B(\cdot,\cdot)$ is the Beta function.
The support of $\phi$ under this distribution is the $[-1,1]$ interval
and stationarity of the log-volatility process is then
guaranteed. \cite{kim98} recommended using $\alpha=20$ and $\beta=1.5$
which implies a prior expectation of $\phi$ equal to 0.86 with a prior
standard deviation of 0.11 thus asigning very little probability mass
for values $\phi<0$. Many authors have chosen to work with this prior
(e.g. \cite{giro11}, \cite{kastner2014ancillarity}, \cite{ehl2016e})
as in practice the parameter $\phi$ is commonly estimated close to 1.
Also, for identifiability the model should be configured with either
$\beta=1$ or $\mu=0$. We chose to work with $\beta=1$ in which case we
asign the prior distribution $\mu\sim\mathcal{N}(\mu_0,\sigma_0^2)$.

Finally, the prior distribution for the parameters $\sigma^2_{\eta}$
was chosen as in \cite{kastner2016Stochvol} for comparison purposes, i.e.
\begin{align}
%\sigma_\eta^2&\sim \mathcal{G}^{-1}\left(\frac{a_\sigma}{2},\frac{b_\sigma}{2}\right),  \label{GI_prior} \\ 
\sigma_\eta^2&\sim B_{\sigma}\times\chi_1^2 = \mathcal{G}\left(\frac{1}{2},\frac{1}{2B_{\sigma}}\right), \label{G_prior}
%\sigma_\eta^2&\sim Inv-\chi^2(c_\sigma,s_\sigma). \label{InvChisq_prior}
\end{align}
also noting that this prior is less influent when the true volatility
of log-volatility is small as it does not bound $\sigma^2_{\eta}$ away
from zero a priori.

We now turn to the prior distributions on the parameter $\nu$ which
depends on the distribution adopted for the error
terms and is assumed independent of $\phi$ and $\sigma^2_{\eta}$. For GED errors we follow \cite{ehl2016e} and 
propose the prior for $\nu\sim$ Inv-$\chi^2$(10,0.05) 
while for
Student-$t$ errors, following \cite{watanabe-asai}, we consider the
truncated exponential density, 
\begin{eqnarray*}
f(\nu) = \lambda \exp\left\{-\lambda(\nu-4)\right\}, ~\nu > 4
\end{eqnarray*}
and zero otherwise, as the prior for $\nu$. However, differently
from \cite{watanabe-asai} we specified $\lambda = 1/3$. For the
Skew-Normal distribution we used $\nu\sim\mathcal{N}(0,5)$.

\subsection{Model Comparison}\label{modelcomparison}

Choosing a model that best represents the data dynamics from a set of
candidate models is challenging in any statistical research.
Despite recent advances in computing the marginal likelihood and the
associated Bayes factor (e.g. \cite{frielp08} and \cite{baur12}), this
remains a difficult and computationaly expensive task in
practice. This motivates us to base model comparison on information
criteria which implementation usually leads to a small extra
computational cost.

A popular choice since the seminal work of \cite{spibcl02} is the
Deviance Information Criterion (DIC). This criterion has been previously 
used for comparing a number of SV models (e.g. \cite{berg-meyer-yu}, 
\cite{abanto-valle-etal}) and is defined as,
\begin{equation*}
\textrm{DIC} = -2\log p(y|\hat{\theta}_{Bayes}) + 2p_{D},
\end{equation*}
where $\theta$ is the vector of parameters in the model, 
$\hat{\theta}_{Bayes} = \mathrm{E}[\theta|y]$ and $p_{D}$ is a penalizing term given by,
\begin{equation*}
p_{D} = 2\Big(\log p(y|\hat{\theta}_{Bayes}) - \mathrm{E}_{\theta|y}[\log p(y|\theta)] \Big).
\end{equation*}
Then, given $S$ simulations from the posterior distribution of $\theta$, $p_D$ 
can be approximated as, 
\begin{equation}
\hat{p}_D = 
2\left(\log p\left(y|\hat{\theta}_{Bayes}\right) - \frac{1}{S}\sum_{s=1}^{S}\log p(y|\theta^s) \right).
\end{equation}

Recent studies however have cautioned against indiscriminate use of DIC as a 
comparison device for latent variable models (e.g. \cite{celeux-etal}, \cite{miller}). 
So, in this paper we propose to compare SV models by looking at the accuracy of 
competing models in predicting out-of-sample observations. This is the idea behind the 
so called Watanabe information criterion (WAIC, \cite{watanabe-2013}, \cite{vehtari-etal}). 
WAIC does not depend on Fisher asymtotic theory and consequently does not assume that the 
posterior distribution converges to a single point, thus providing an attractive alternative
to compare hierarchical models. It can be interpreted as a computationally convenient 
approximation to cross-validation and is defined as,
\begin{equation}\label{elpd}
\textrm{elpd}_{waic} = \textrm{lpd} - p_{waic},
\end{equation}
where $p_{waic}=\sum_{i=1}^{n}\mathrm{Var}_{\theta|y}(\log p(y_i|\theta))$ penalizes for the 
effective number of parameters
and $\textrm{lpd} = \sum_{i=1}^{n}\log p(y_i|y)$ is the log pointwise predictive density 
as defined in \cite{vehtari-etal} with each term in the sum given by,
\begin{equation*}\label{logp}
\log p(y_i|y) = \log \int_{\Theta}p(y_i|\theta)p(\theta|y)d\theta. 
\end{equation*}

\noindent In practice, given $S$ simulated values of $\theta$ from its posterior distribution, the two terms in (\ref{elpd}) are estimated as,
\begin{align*}\label{lpdhat}
\widehat{\textrm{lpd}} = \sum_{i=1}^{n}\log\left(\frac{1}{S}\sum_{s=1}^{S}p(y_i|\theta^s)\right). 
\end{align*}  

\noindent and
\begin{align*}
\hat{p}_{waic}  = \sum_{i=1}^{n}V_{s=1}^S\big(\log p(y_i|\theta)\big).
\end{align*}  
where $V_{s=1}^S$ denotes the sample variances of
$\log p(y_i|\bftheta^{(1)}),\dots,\log p(y_i|\bftheta^{(S)})$,
$i=1,\dots,n$.
The WAIC is then defined as,
\begin{equation*}
\textrm{WAIC} = \widehat{\textrm{lpd}}- \hat{p}_{waic},
\end{equation*}
or equivalently as $-2\widehat{\textrm{lpd}} +2\hat{p}_{waic}$ to be on the deviance scale.
\cite{waic} showed that under certain regularity conditions, the WAIC is asymptotically 
equivalent to leave-one-out cross-validation (see also, \cite{gelman2014}).

The log pointwise predictive density can also be estimated via approximate leave-one-out cross-validation (LOO) as,
\begin{equation*}
\textrm{lpd}_{loo} = \sum_{i=1}^{n}\log p(y_i|y_{-i}) = \sum_{i=1}^{n}\log \int_{\Theta}p(y_i|\theta)p(\theta|y_{-i})d\theta,
\end{equation*}
where $y_{-i}$ denotes the data vector with the $i$th observation deleted.
\cite{vehtari2017practical} introduced an efficient approach to compute LOO using Pareto-smoothed
importance sampling (PSIS) for regularizing importance weights. The computations are implemented in the {\tt R} 
packaged {\tt loo} which we use here together with Stan to perform model comparison. An interesting byproduct of this approach is that
approximate standard errors for estimated predictive errors are also obtained.

\section{Simulations} \label{simulations}

In this section, a simulation study is carried out to compare the two
sampling approaches in {\tt stochvol} and Stan to estimate SV models with
normal errors. We generated $m=100$ replications of time series with
$500$, $1000$ and $1500$ observations from the model described in
(\ref{modSV})-(\ref{log_vol}) with $\beta=1$ for identifiability. The model parameters were
fixed as follows, $\mu = -9$, $\phi \in \{0.95; 0.99\}$ and $\sigma_\eta \in \{0.05; 0.15\}$. 

The prior distributions assumed were as described in (\ref{priorPhi})
for the parameter $\phi$, $\mu\sim\mathcal{N}(-10,1)$ and the prior in
(\ref{G_prior}) was asigned for $\sigma^2_\eta$ with $B_\sigma=0.1$. These were the
prior distributions used in \cite{kastner2016Stochvol} to describe the
{\tt stochvol} package and are adopted here in both
sampling methods for comparison purposes. Then, for each series we simulated 10,000 MCMC
samples with the first 5,000 discarded as burn-in. The estimation
performance was evaluated considering two criteria, the bias and the
square root of the mean square error (smse), defined as,
\begin{align*}
\textrm{bias} = \frac{1}{m}\sum_{i=1}^{m}\hat{\theta}^{(i)} - \theta, \qquad
\textrm{smse}^2 = \frac{1}{m}\sum_{i=1}^{m}(\hat{\theta}^{(i)} - \theta)^2
\end{align*}
where $\hat{\theta}^{(i)}$ is the point estimate of parameter $\theta$
in the $i$-th replication, $i=1,\ldots,m$. 

The results for $\mu=-9$ and combinations of $\phi$ and $\sigma_{\eta}$ are shown in Tables
\ref{tab:sim1}, \ref{tab:sim2}, \ref{tab:sim3} and \ref{tab:sim4}. The
first two tables refer to a very high degree of persistence in the
log-volatility process ($\phi=0.99$). 
In Table \ref{tab:sim1} for $\sigma_\eta=0.15$, we obtained good
results in terms of bias and smse for all parameters with HMC doing
better for larger sample sizes. For series with smaller sizes
($n=500$) the results are close for both methods (slightly better for
{\tt stochvol}) except for the 
parameter $\mu$ which shows a much smaller bias when using {\tt stochvol}.
In Table \ref{tab:sim2} ($\sigma_\eta=0.05$), we notice much smaller
values for bias and smse for both methods, but with HMC doing better overall.

\begin{center} [ Table \ref{tab:sim1} around here ] \end{center}

\begin{center} [ Table \ref{tab:sim2} around here ] \end{center}

Tables \ref{tab:sim3} and \ref{tab:sim4} refer to a log-volatility
process with a smaller persistence ($\phi=0.95$). From the
results in these tables we first notice that both methods provide better estimates
compared to the case with higher persistence and this is in general
expected. However, for all sample sizes and parameters the HMC method
using Stan is doing even better compared to {\tt stochvol}. Also, the resulting
Markov chains using HMC (not shown to save space) in general present
lower autocorrelations thus mixing better.

\begin{center} [ Table \ref{tab:sim3} around here ] \end{center}

\begin{center} [ Table \ref{tab:sim4} around here ] \end{center}

\section{Applications} \label{applications}

In this section we illustrate the use of HMC methods via Stan for the
Bayesian estimation of SV models
with different distributions for the observation error $\epsilon_t$. The illustration uses two exchange
rate time series data: the Pound/Dollar (\pounds/USD) and the
Euro/Dollar (EUR/USD), plus the stock index in S\~ao Paulo
(IBOVESPA). The time series are the daily continuously compounded returns in 
percentage, defined as $y_t = 100[\log(P_t)-\log(P_{t-1})]$ where
$P_t$ is the price at time $t$. The \pounds/USD series of daily
returns in percentage covers the period from 1/10/81 to
28/6/85 and was studied before by for example
\cite{har94} and \cite{ehl2016e}. The series EUR/USD covers the
period from 03/01/2004 to 04/04/2012 and is available in the {\tt
  stochvol} package. Finally, the series IBOVESPA covers the period
from 03/01/2005 to 28/02/2013. 

\begin{figure}[h]\centering
\includegraphics{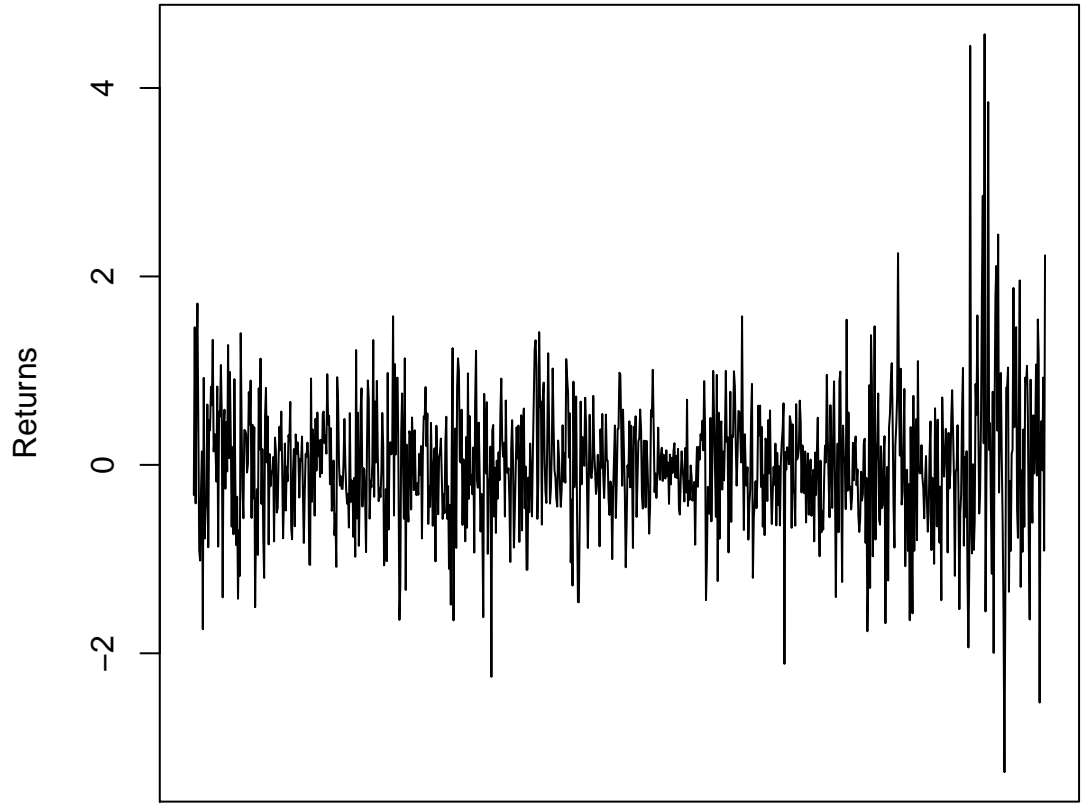}
\caption{Pound/Dollar time series returns.}
\label{fig:SeriesA1}
\end{figure}

\begin{figure}[h]\centering
\includegraphics{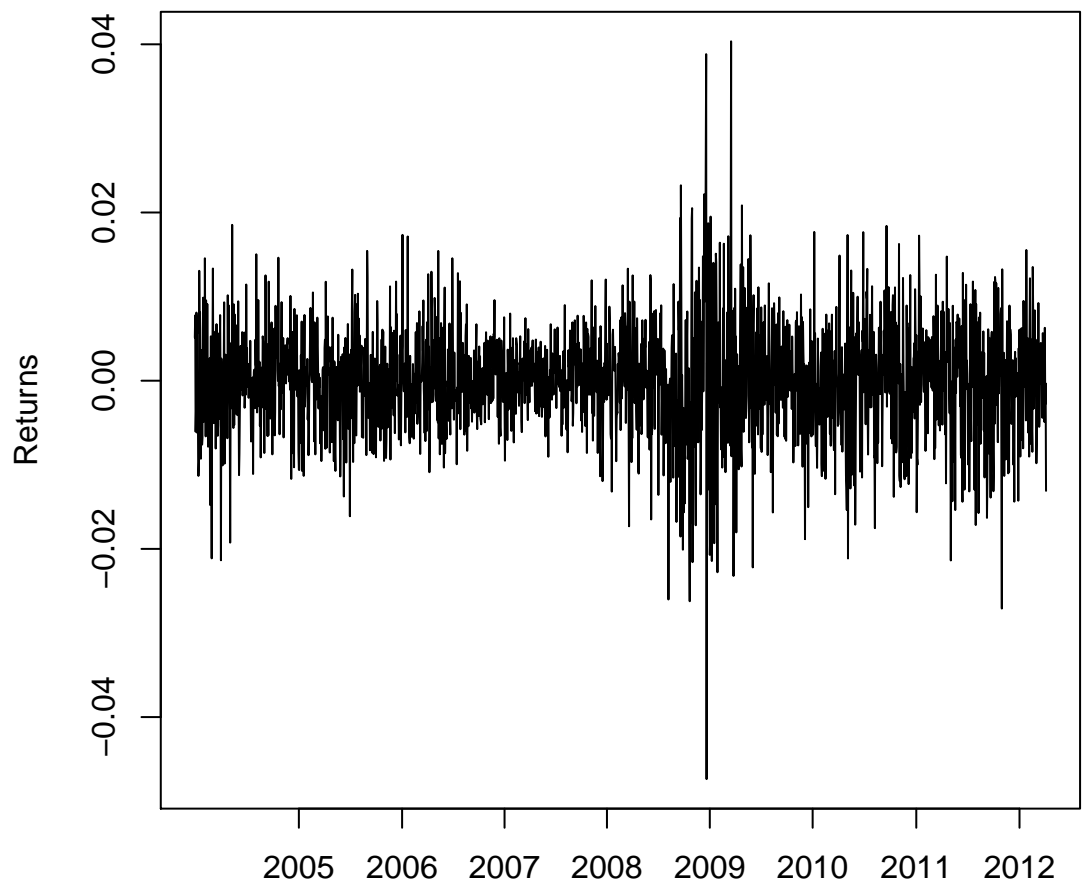}
\caption{Euro/Dollar time series returns.}\label{fig:SeriesA2}
\end{figure}

\begin{figure}[h]\centering
\includegraphics{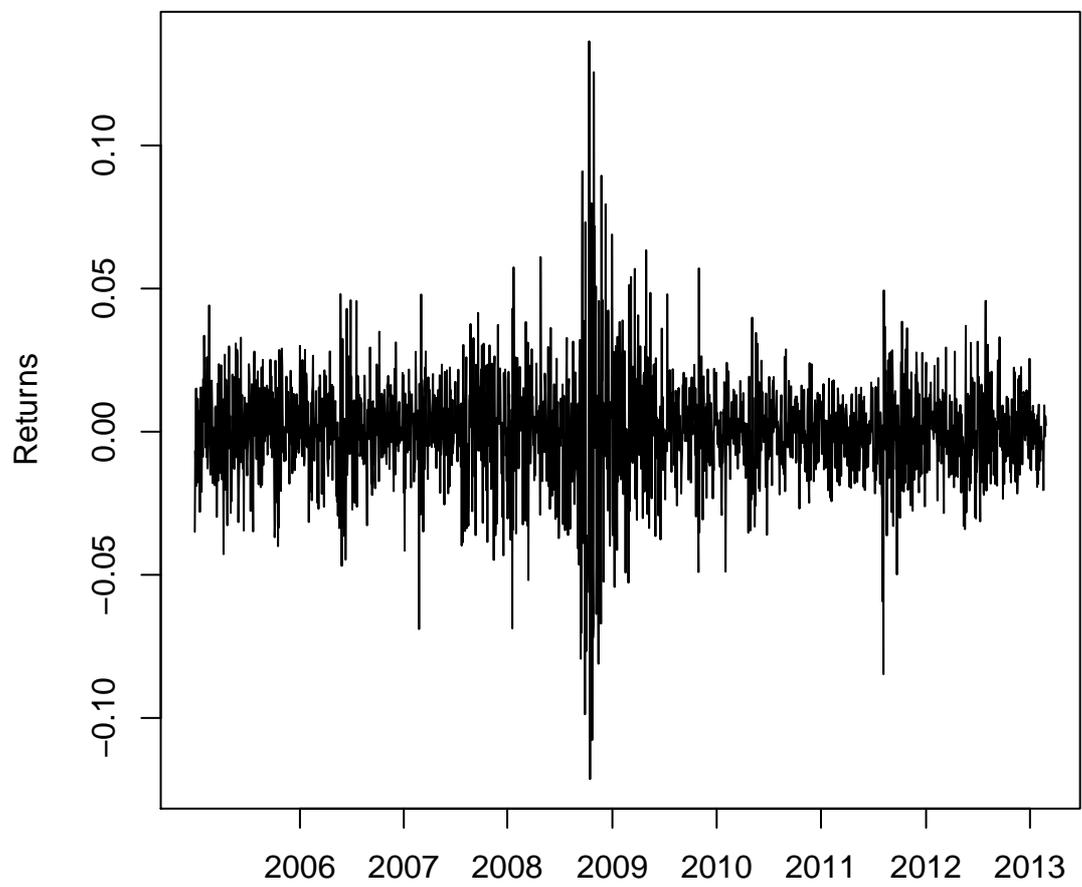}
\caption{IBOVESPA time series returns.}\label{fig:SeriesA3}
\end{figure}

The \pounds/USD, EUR/USD and IBOVESPA time series have 945, 2120 and
2016 observations respectively. The series are depicted in Figures
\ref{fig:SeriesA1}, \ref{fig:SeriesA2} and \ref{fig:SeriesA3} and
Table \ref{tab:summary1} consigns some descriptive statistics for
these series. From this table we notice high kurtosis for all series
and a little skewness for the two exchange rate series. Finally, the
autocorrelation function (not shown) indicated no serial correlation in the three
series. 
 
\begin{center} [ Table \ref{tab:summary1} around here ] \end{center}

\begin{center} [ Figure \ref{fig:SeriesA1} around here ] \end{center}

\begin{center} [ Figure \ref{fig:SeriesA2} around here ] \end{center}

\begin{center} [ Figure \ref{fig:SeriesA3} around here ] \end{center}

For each time series, the SV model was estimated using demeaned
returns and considering the four different distributions for the errors:
the Gaussian, the GED distribution with parameter $\nu$, the   
Student's $t$ distribution with $\nu$ degrees of freedom and the
Skew-Normal distribution. We considered prior distributions as
described in (\ref{priorPhi}) for the parameter $\phi$,
$\mu\sim\mathcal{N}(-10,1)$ and the Gamma prior (\ref{G_prior}) was
asigned for $\sigma_\eta$ with
hyperparameter $B_\sigma=0.1$. For the parameter $\nu$ in the GED,
$t$-Student and Skew-Normal the prior distributions are
Inv-$\chi^2$(10,0.05), truncated exponential and
$\nu\sim\mathcal{N}(0,5)$ respectively as
described in the end of Section \ref{methods}. 

Then, for each time series we drew 10,000 MCMC samples
for the parameters and volatilities using Stan and NUTS where the
first 5,000 were discarded as burn-in. To compare models with
different error distributions we used the information criteria
described in Section (\ref{CritSel}). The results are reported in Tables
\ref{tab:CompMod} and \ref{tab:EstA}.

According to Table \ref{tab:CompMod}, we can see that for the
\pounds/USD series the information criteria have similar values for
the Gaussian and Skew-Normal distributions. DIC and LOO show slightly
lower values for the Gaussian model although our descriptive analysis
has shown high kurtosis and slight skewness for this series.
For the EUR/USD returns the SV model with $t$-Student and
GED errors show better results in terms of all criteria with DIC selecting
the $t$-Student while WAIC and LOO have lower values for the
GED. Finally, for the IBOVESPA series all criteria indicate Gaussian
and Skew-Normal errors as most appropriate SV models.

\begin{center} [ Table \ref{tab:CompMod} around here ] \end{center}

\begin{center} [ Table \ref{tab:EstA} around here ] \end{center}

The parameter estimates (posterior mean and standard deviation)
obtained for the three series are shown in \ref{tab:EstA}. We notice
that the standard deviation of the parameter $\mu$ for the EUR/USD and
IBOVESPA are quite smaller than for the \pounds/USD series and the
persistence $\phi$ is quite high for all series indicating that the
volatility in the previous day tends to have an impact on the current exchange
rates prices.
We also notice that estimates for the parameter $\nu$ in the
Skew-Normal distribution were close to zero for the three series with
0.95 credible intervals given by $(-0.123; 0.045)$, $(-0.036; 0.072)$
and $(-0.0334; 0.0764)$. However, for the
\pounds/USD and IBOVESPA series this models presented good results in
terms of information criteria.

\subsection{Sensitivity to the Choice of Prior}

In this section we investigate whether the prior choice for the
parameter $\sigma_{\eta}$ can influence the results for model
comparison. We estimated the 
SV model for the \pounds/USD series considering the prior
distributions described in (\ref{GI_prior}), (\ref{G_prior1}) and
(\ref{InvChisq_prior}) below,

\begin{align}
\sigma_\eta^2&\sim \mathcal{G}^{-1}\left(\frac{a_\sigma}{2},\frac{b_\sigma}{2}\right),  \label{GI_prior} \\ 
\sigma_\eta^2&\sim B_{\sigma}\times\chi_1^2 = \mathcal{G}\left(\frac{1}{2},\frac{1}{2B_{\sigma}}\right), \label{G_prior1} \\ 
\sigma_\eta^2&\sim Inv-\chi^2(c_\sigma,s_\sigma). \label{InvChisq_prior}
\end{align}

The sensitivity was evaluated considering the information criteria for
model comparison and checking if the model rankings changed with the
choice of prior. In this estimation we simulated 5,000 samples from
the posterior distribution using Stan from which 50\% were discarded
as burn-in resulting in a final sample of 2500 values. The four
distributions for the error term in the SV model were considered and
we repeated this procedure twice.

The results appear in Table \ref{tab:Sens} where we notice that the
choice of the prior distribution for $\sigma_{\eta}$ does not seem to
influence model selection for this series. So, according to these
criteria the Gaussian and Skew-Normal models are still prefered. It is
worth noting that even with only 2500 samples used in the computations
the HMC method converged rather quickly and the values of the criteria
were similar to the ones obtained with 5000 effective samples (after burn-in).

\begin{center} [ Table \ref{tab:Sens} around here ] \end{center}

\section{Conclusions}\label{conclusion}
  
In this paper we discuss and compare the Bayesian estimation approach in stochastic volatility models with heavy tailed and possibly asymmetric distributions for the error term. We employed both traditional Markov chain Monte Carlo and Hamiltonian Monte Carlo methods to obtain approximations to the posterior marginal distributions of interest.

In particular we compared very fast algorithms which implementation is freely available in {\tt R} packages, namely the {\tt stochvol} package and the {\tt rstan} package (an interface with the Stan package).
These methods were assessed through a simulation study and the Stan package was also tested with real time series of returns. Overall, we found evidence that Stan is slightly more efficient for estimation with the advantage of being able to deal with different model structures and distributions.
We hope that our findings are useful to the practitioners.

\section*{Acknowledgments}

Ricardo Ehlers received
support from S\~ao Paulo Research Foundation (FAPESP) - Brazil, under
grant number 2016/21137-2.

%\bibliographystyle{plainnat}
%\bibliography{mybib,mypub}

\clearpage

\begin{table}[h]
\begin{center}     
\caption{Bias and square root of mean square error (smse) for parameter estimates.
  True parameters: $\mu=-9$, $\phi=0.99$ and  $\sigma_\eta=0.15$.} 
\label{tab:sim1}\vskip .3cm
\begin{tabular}{cccccccc}\hline \hline 
\multirow{2}{*}{T}& \multirow{2}{*}{Method}&\multicolumn{2}{c}{$\mu$}&
\multicolumn{2}{c}{$\phi$} &
\multicolumn{2}{c}{$\sigma_\eta$}\\ \cline{3-8} 
                     &           & Bias        &smse & Bias        & smse& Bias        &smse \\ \hline
\multirow{2}{*}{500} &	Stochvol &  0.213      & 0.474 & 0.020       & 0.030 & -0.137      & 0.121 \\
		     & HMC  	 &  0.491      & 0.491 & 0.068       & 0.066 & -0.117      & 0.118 \\ \cline{2-8}
\multirow{2}{*}{1000}&	Stochvol &  0.223      & 0.415 & 0.007       & 0.011 & -0.019      & 0.031 \\ 
		     & HMC  	 &  0.055      & 0.056 & 0.002       & 0.002 & -0.006      & 0.007 \\ \cline{2-8} 
\multirow{2}{*}{1500}&	Stochvol &  0.129      & 0.366 & 0.004       & 0.007 & -0.012      & 0.024 \\
		     & HMC  	 &  0.120      & 0.267 & 0.003       & 0.007 & -0.007      & 0.021 \\       
\hline\hline					  
\end{tabular}
\end{center}
\end{table}

\clearpage

\begin{table}[h]
\begin{center}     
\caption{Bias and square root of mean square error (smse) for parameter estimates.
  True parameters: $\mu=-9$, $\phi=0.99$ and $\sigma_\eta=0.05$.} 
\label{tab:sim2}\vskip .3cm
\begin{tabular}{cccccccc}\hline \hline 
\multirow{2}{*}{T}& \multirow{2}{*}{Method}&\multicolumn{2}{c}{$\mu$}&
\multicolumn{2}{c}{$\phi$} &
\multicolumn{2}{c}{$\sigma_\eta$}\\ \cline{3-8} 
		     &           & Bias        &smse & Bias        & smse& Bias        &smse \\ \hline
\multirow{2}{*}{500} &	Stochvol &  0.063      & 0.218 & 0.106       & 0.121 & -0.071      & 0.082 \\
		 &  HMC  	 &  0.090      & 0.093 & 0.132       & 0.133 &  0.023      & 0.029 \\ \cline{2-8}
\multirow{2}{*}{1000}&	Stochvol &  0.041      & 0.146 & 0.042       & 0.064 & -0.039      & 0.049 \\
		 &  HMC  	 &  0.018      & 0.134 &-0.001       & 0.002 & -0.010      & 0.012 \\ \cline{2-8} 
\multirow{2}{*}{1500}&	Stochvol &  0.017      & 0.122 & 0.027       & 0.047 & -0.027      & 0.037 \\
		 & HMC  	 &  0.017      & 0.123 & 0.018       & 0.034 & -0.015      & 0.026 \\         
\hline\hline					  			  
\end{tabular}
\end{center}
\end{table}

\clearpage

\begin{table}[h]
\begin{center}     
\caption{Bias and square root of mean square error (smse) for parameter estimates.
  True parameters: $\mu=-9$, $\phi=0.95$ and $\sigma_\eta=0.15$.} 
\label{tab:sim3}\vskip .3cm
\begin{tabular}{cccccccc}\hline \hline 
\multirow{2}{*}{T}& \multirow{2}{*}{Method}&\multicolumn{2}{c}{$\mu$}&
\multicolumn{2}{c}{$\phi$} &
\multicolumn{2}{c}{$\sigma_\eta$}\\ \cline{3-8} 
		     &           & Bias        &smse & Bias        & smse& Bias        &smse \\ \hline
\multirow{2}{*}{500} &	Stochvol &  0.049      & 0.151 & 0.050       & 0.066 & -0.042      & 0.059 \\ 
                     & HMC       &  0.042      & 0.150 & 0.030       & 0.049 &  0.000      & 0.051 \\ \cline{2-8}
\multirow{2}{*}{1000}&	Stochvol &  0.022      & 0.107 & 0.025       & 0.045 & -0.022      & 0.044 \\ 
		 & HMC  	 &  0.019      & 0.106 & 0.014       & 0.035 & -0.003      & 0.038 \\ \cline{2-8} 
\multirow{2}{*}{1500}&	Stochvol &  0.027      & 0.083 & 0.018       & 0.030 & -0.023      & 0.042 \\
		 & HMC  	 &  0.024      & 0.082 & 0.011       & 0.024 & -0.011      & 0.035 \\       
\hline\hline					  
\end{tabular}
\end{center}
\end{table} 

\clearpage

\begin{table}[h]
\begin{center}     
\caption{Bias and square root of mean square error (smse) for parameter estimates.
  True parameters: $\mu=-9$, $\phi=0.95$ and$\sigma_\eta=0.05$.} 
\label{tab:sim4}
\vskip .3cm
\begin{tabular}{cccccccc}\hline \hline 
\multirow{2}{*}{T}& \multirow{2}{*}{Method}&\multicolumn{2}{c}{$\mu$}&
\multicolumn{2}{c}{$\phi$} &
\multicolumn{2}{c}{$\sigma_\eta$}\\ \cline{3-8} 
		     &           & Bias        &smse & Bias        & smse  & Bias        &smse \\ \hline
\multirow{2}{*}{500} &	Stochvol &  0.025      & 0.079 & 0.119     & 0.124 & -0.050      & 0.060 \\
		     &  HMC  	 &  0.011      & 0.076 & 0.086     & 0.091 &  0.009      & 0.037 \\ \cline{2-8}
\multirow{2}{*}{1000}&	Stochvol &  0.008      & 0.059 & 0.118     & 0.123 & -0.046      & 0.060 \\
		 &  HMC  	 & -0.001      & 0.057 & 0.087     & 0.092 &  0.001      & 0.043 \\ \cline{2-8} 
\multirow{2}{*}{1500}&	Stochvol &  0.005      & 0.042 & 0.116     & 0.122 & -0.037      & 0.049 \\
		 & HMC  	 & -0.001      & 0.044 & 0.087     & 0.097 & -0.006      & 0.037 \\         
\hline\hline					  			  
\end{tabular}
\end{center}
\end{table}

\clearpage

\begin{table}[h]
\begin{center}     
\caption{Descriptive statistics of returns under study, $T$ is the number of observations.}\label{tab:summary1} 
\begin{tabular}{lrcccc}\hline\hline 
Series      & $T$  & Mean    & Std Dev & Skewness   & Kurtosis\\ \hline
\pounds/USD & 945  &-0.03530 & 0.7111  &  0.60      & 7.85    \\
EUR/USD     & 2120 & 0.00001 & 0.0066  & -0.15      & 6.18    \\  
IBOVESPA    & 2016 & 0.00041 & 0.0188  & -0.04      & 8.89    \\    
\hline\hline					  
\end{tabular}
\end{center}
\end{table}

\clearpage

\begin{table}[h]
\begin{center}     
\caption{Comparison between the proposed models via information criteria.}\label{tab:CompMod} 
\begin{tabular}{llccccc}\hline\hline 
Series		         & Dist.       &	DIC&	WAIC	& SE$_{waic}$ &	LOO 	&	SE$_{loo}$ \\\hline
\multirow{4}{*}{\pounds/USD}& Gaussian &   1802.7	&   1807.4   &	53.5	&	1811.8&	54.3   \\
	    	         &  t-Student  &   1805.3	&   1812.8   &	52.8	&	1813.5&	52.9   \\
			 &  Skew-Normal&   1803.1	&   1807.4   &	53.3	&	1812.2&	54.2   \\
			 &  GED	       &   1810.6	&   1811.3   &	53.2	&	1814.0&	53.7   \\\cline{2-7}
\multirow{4}{*}{EUR/USD} &  Gaussian   & -15647.9	& -15640.2   &	77.0	&-15638.2 &	77.3   \\
			 &  t-Student  & -15664.9	& -15647.8   &	76.4	&-15647.4 &	76.4   \\
			 &  Skew-Normal& -15646.7	& -15638.9   &	77.2	&-15636.6 &	77.5   \\
			 &  GED	       & -15651.2	& -15650.9   &	76.6	&-15650.2 &	76.7   \\\cline{2-7}
\multirow{4}{*}{IBOVESPA}&  Gaussian   & -10986.7	& -10983.7   &	74.9	&-10975.9 &	75.6   \\
			 &  t-Student  & -10974.6 	& -10970.3   &	75.7	&-10968.0 &	75.9   \\
			 &  Skew-Normal& -10984.3	& -10981.0   &	75.1	&-10973.7 &	75.8   \\
			 &  GED	       & -10977.3	& -10979.1   &	75.4	&-10974.4 &	75.8   \\      
\hline\hline
\end{tabular}
\end{center}
\end{table}

\clearpage

\begin{table}[h]
\begin{center}     
\caption{Parameter estimates, posterior mean and standard deviations
  (in parenthesis)}.\label{tab:EstA} 
\begin{tabular}{llcccc}\hline\hline 
Series			 & Dist. &	$\mu$	     	    &	$\phi$	      &$\sigma_\eta$   &	$\nu$   \\ \hline
\multirow{4}{*}{\pounds/USD}& Gaussian & - 9.580 (1.014)  & 0.999 (0.000) & 0.137 (0.024)     &    	        \\
			 & t-Student   & - 9.591 (1.046)  & 0.999 (0.000) & 0.113 (0.024)     &12.143 (3.581)	\\
			 & Skew-Normal & - 9.606 (1.026)  & 0.999 (0.000) & 0.141 (0.026)     &-0.038 (0.042)	\\
			 & GED         & - 9.592 (1.043)  & 0.999 (0.000) & 0.123 (0.025)     & 1.763 (0.141)	\\ \cline{2-6}
\multirow{4}{*}{EUR/USD} & Gaussian    & -10.197 (0.288)  & 0.994 (0.002) &	0.065 (0.010) &    	        \\
			 & t-Student   & -10.321 (0.289)  & 0.994 (0.002) &	0.059 (0.010) &14.090 (3.393)	\\
			 & Skew-Normal & -10.199 (0.290)  & 0.994 (0.002) &	0.065 (0.010) & 0.016 (0.027)	\\
			 & GED         & -10.175 (0.312)  & 0.995 (0.002) &	0.055 (0.010) & 1.681 (0.082)	\\\cline{2-6}
\multirow{4}{*}{IBOVESPA}& Gaussian    & - 8.416 (0.199)  & 0.980 (0.006) &	0.148 (0.019) &    	        \\
			 & t-Student   & - 8.546 (0.236)  & 0.984 (0.005) &	0.128 (0.017) &16.202 (4.229)	\\
			 & Skew-Normal & - 8.416 (0.211)  & 0.981 (0.006) &	0.145 (0.020) & 0.020 (0.028)	\\
			 & GED         & - 8.417 (0.216)  & 0.983 (0.005) &	0.132 (0.018) & 1.801 (0.107)	\\
\hline\hline
\end{tabular}
\end{center}
\end{table}

\clearpage

\begin{table}[h]
\begin{center}
\caption{Comparison of proposed models via information criteria for the sensitivity study.}\label{tab:Sens} 
\begin{tabular}{llccccc}
\hline\hline
Prior for $\sigma_\eta$	 & Dist.        &	DIC     &	WAIC	& EP$_{waic}$ &	LOO 	&	EP$_{loo}$ \\\hline
\multirow{8}{*}{$\mathcal{G}^{-1}(2.5, 0.025)$}
& \multirow{2}{*}{Gaussian}&   1805.0	&	1810.3	&	53.8	&	1811.0	&	54.5	\\
&			   &   1801.4	&	1805.8	&	53.4	&	1809.6	&	54.0	\\
& \multirow{2}{*}{$t$-Student}&   1804.5	&	1811.5	&	52.8	&	1812.3	&	52.9	\\
&			  &   1805.7	&	1813.4	&	52.8	&	1814.2	&	52.9	\\
& \multirow{2}{*}{Skew-Normal}&   1803.7	&	1808.3	&	53.4	&	1812.0	&	54.0	\\
&			  &   1803.4	&	1807.7	&	53.3	&	1811.4	&	53.9	\\
& \multirow{2}{*}{GED}	  &   1809.7	&	1810.4	&	53.2	&	1813.2	&	53.0	\\
& 			  &   1812.6	&	1813.3	&	53.2	&	1815.5	&	53.6	\\
\cline{2-7}													 	 
\multirow{8}{*}{$\mathcal{G}\left(\frac{1}{2},5\right)$} 
& \multirow{2}{*}{Gaussian}  &	1800.9	&	1805.0	&	53.1	&	1809.3	&	53.8	\\
&			  &	1801.7	&	1806.8	&	53.4	&	1811.1	&	54.1	\\
& \multirow{2}{*}{$t$-Student}  &	1805.3	&	1812.8	&	52.8	&	1813.4	&	52.9	\\
&			  &	1804.0	&	1810.7	&	52.9	&	1811.5	&	52.9	\\
& \multirow{2}{*}{Skew-Normal}&	1804.0	&	1808.8	&	53.5	&	1812.7	&	54.2	\\
&			  &	1803.9	&	1808.2	&	53.3	&	1812.1	&	54.0	\\
& \multirow{2}{*}{GED}	  &	1809.0	&	1809.5	&	53.0	&	1812.0	&	53.4	\\
&			  &	1811.4	&	1812.2	&	53.2	&	1815.0	&	53.7	\\\cline{2-7}	
\multirow{8}{*}{$Inv-\chi^2(10. 0.05)$}    	
& \multirow{2}{*}{Gaussian}  &	1803.4	&	1808.0	&	53.4	&	1811.7	&	54.0	\\
&			  &	1804.2	&	1809.0	&	53.5	&	1813.0	&	54.3	\\
& \multirow{2}{*}{$t$-Student}  &	1805.8	&	1813.5	&	52.8	&	1814.1	&	52.9	\\
&			  &	1807.3	&	1815.1	&	52.9	&	1815.7	&	53.0	\\
& \multirow{2}{*}{Skew-Normal}&	1804.2	&	1808.5	&	53.4	&	1811.6	&	53.9	\\
&			  &	1802.9	&	1807.6	&	53.3	&	1811.2	&	54.0	\\
& \multirow{2}{*}{GED}	  &	1811.3	&	1812.0	&	53.2	&	1813.9	&	53.5	\\
&			  &	1814.8	&	1815.5	&	53.3	&	1817.0	&	53.6	\\
\hline\hline					  
\end{tabular}
\end{center}
\end{table}	

\clearpage

\clearpage

\clearpage

\end{document}